# Long-wave mid-infrared cavity-enhanced frequency comb spectroscopy of cold, complex molecules

NEGAR BARADARAN,[1,†] DOMINIK CHARCZUN,[1,†] TANAY NAMBIAR,[1] MEIJUN ZOU,[1] KEVIN F. LEE,[2] MARTIN E. FERMANN,[2] AND MARISSA L. WEICHMAN[1,*]

[1]*Department of Chemistry, Princeton University, Princeton, New Jersey, 08544, USA*
[2]*IMRA America Inc., 1044 Woodridge Ave, Ann Arbor, Michigan 48105, USA*
[†]*These authors contributed equally.*
[*]*weichman@princeton.edu*

**Abstract:** We report the development of a new instrument for cavity-enhanced absorption spectroscopy of the fundamental rovibrational transitions of molecules in the long-wave mid-infrared (LWIR) region from 6500 to 10000 nm. Our setup combines a LWIR frequency comb, a high-finesse optical enhancement cavity, a cryogenic buffer gas cooling cell, and a Fourier transform interferometer to obtain broadband, high-resolution, and high-sensitivity molecular absorption spectra. Here, we showcase the capabilities of this setup by presenting the gas-phase LWIR spectra of the $\nu_6$ band of ethane and the $\nu_{10}$ band of the *gauche* conformer of ethanol near 7200 nm under both room temperature and cryogenic conditions.

## 1. Introduction

Interpretation and assignment of molecular spectra in complex environments is driven by laboratory spectroscopy of known species. Hydrocarbons in particular play key roles in astrochemistry [1–4], combustion chemistry [5,6], and environmental analysis [7,8]. Despite experimental advances, the spectra of many complex hydrocarbons remain poorly characterized due to challenges including (a) closely-spaced, narrow spectral features that require high spectral resolution to resolve; (b) weak absorption cross sections and low accessible number densities which necessitate high-sensitivity detection; and (c) dense manifolds of rovibrational states that are best captured with broadband spectroscopy. Cavity-enhanced frequency comb spectroscopy (CE-FCS) is a powerful technique that can meet all these requirements – high spectral resolution, high detection sensitivity, and broadband acquisition – simultaneously.

Frequency comb lasers were originally developed for frequency metrology [9–11] and are now among the most powerful tools at our disposal for direct molecular spectroscopy [12,13]. Frequency combs output broad spectral envelopes consisting of sharp, evenly spaced laser lines or "teeth" with optical frequencies given by $\nu_m = m \times f_{\mathrm{rep}} + f_0$ where $m$ is an integer, $f_{\mathrm{rep}}$ is the comb repetition rate, and $f_0$ is the carrier-envelope offset frequency. The $\nu_m$ frequencies can be stabilized with high precision and accuracy through active locking of $f_{\mathrm{rep}}$ and $f_0$. Combs are therefore simultaneously broadband and precise light sources acting as an array of stable individual continuous-wave (cw) lasers lasing synchronously. In CE-FCS, the periodic comb structure is coupled into the longitudinal modes of an optical enhancement cavity containing the absorber of interest. Such coupling is achieved by matching integer multiples of the free spectral range ($FSR$) of the enhancement cavity and the comb $f_{\mathrm{rep}}$. Cavity coupling enhances the interaction length between light and sample and can thereby increase absorption sensitivity by a factor proportional to the cavity finesse, $F$, which is often several orders of magnitude.

CE-FCS is particularly powerful in combination with cryogenic cooling to reduce the molecular partition function and alleviate spectral congestion. Molecules with as few as 10 atoms can occupy millions of vibrational and rotational states at room temperature, yielding congested spectra that are difficult to resolve, interpret, and assign. By cooling the ensemble's translational and internal degrees of freedom to cryogenic temperatures, one can drastically narrow Doppler linewidths and condense molecular population in a small collection of rotational levels of the vibrational ground state. Cold conditions therefore allow one to take full advantage of the frequency resolution afforded by comb spectroscopy. Nearly any molecule can be cooled to temperatures approaching absolute zero through collisions with a cold, inert gas inside a cryogenic buffer gas cell (CBGC) [14–17]. CE-FCS combined with cryogenic cooling paves the way towards the spectroscopy of a wide array of species with frequency precision unattainable with other broadband techniques.

Here, we present a CE-FCS setup operating in the long-wave mid-infrared (LWIR) integrated with a 4 K helium-cooled CBGC and Fourier transform spectrometer (FTS) to enable broadband, high-resolution, and high-sensitivity rovibrational spectroscopy of cold gas-phase molecules. We make use of difference frequency generation (DFG) to produce LWIR comb light spanning 6500−10000 nm, with an instantaneous spectral bandwidth on the order of 1000 nm. Working in the LWIR is experimentally convenient for characterization and detection of spectroscopically challenging species: fundamental low-frequency vibrational transitions are accessible, Doppler broadened linewidths are reduced, and the density of nearby molecular dark states is minimized at low internal energies. This last point is key to limit intramolecular vibrational energy redistribution (IVR) and thereby avoid intrinsic spectral congestion. On the other hand, spectroscopy in the LWIR can also permit careful study of the onset of IVR, state mixing, and intrinsic congestion in molecular systems of increasing complexity. Here, we report fully-resolved direct absorption spectra of cold gas-phase ethane and ethanol to demonstrate the capabilities of our LWIR CE-FCS instrument. The data we report here represent, to the best of our knowledge, the first high-resolution spectrum of ethane at cryogenic temperatures and the first rovibrationally-resolved spectrum of ethanol near 7200 nm.

## 2. Methods

Our spectroscopic apparatus is comprised of four main components: the frequency comb light source, the buffer gas cooling system, the optical enhancement cavity, and the Fourier transform spectrometer. In this section we provide a detailed description of each of these systems. We also provide the step-by-step procedure by which we obtain comb-mode-resolved FTS spectra that surpass the path-length resolution limit.

### 2.1 LWIR frequency comb source

Our frequency comb is based on an Er:fiber oscillator that generates pulses centered at 1550 nm with $f_{rep}$ near 93.4 MHz [18]. A small portion of the oscillator output is picked off and sent to a fast photodiode (EOT ET-3000A) to monitor the comb repetition rate. The main oscillator output is split into power and signal arms to ultimately enable DFG and produce LWIR light (Fig. 1a) [19]. The power arm contains a two-stage Er:fiber amplifier yielding 3.5 W of light (38 nJ/pulse). Pulse shaping by an array of resistive heaters on the fiber Bragg grating stretcher adjusts dispersion for better compression, achieving a pulse duration of 90 fs [20]. In parallel, the signal arm is pre-amplified in Er:fiber then broadened in a highly nonlinear fiber to generate a supercontinuum spanning 600−2000 nm. The signal arm is then amplified in Tm:fiber, yielding 600 mW of light with a pulse duration of 60 fs. We set the polarization in each arm using a fiber polarization controller. We additionally set the intensity of the power arm using a half-waveplate and polarizing beam splitter after the light exits the fiber.

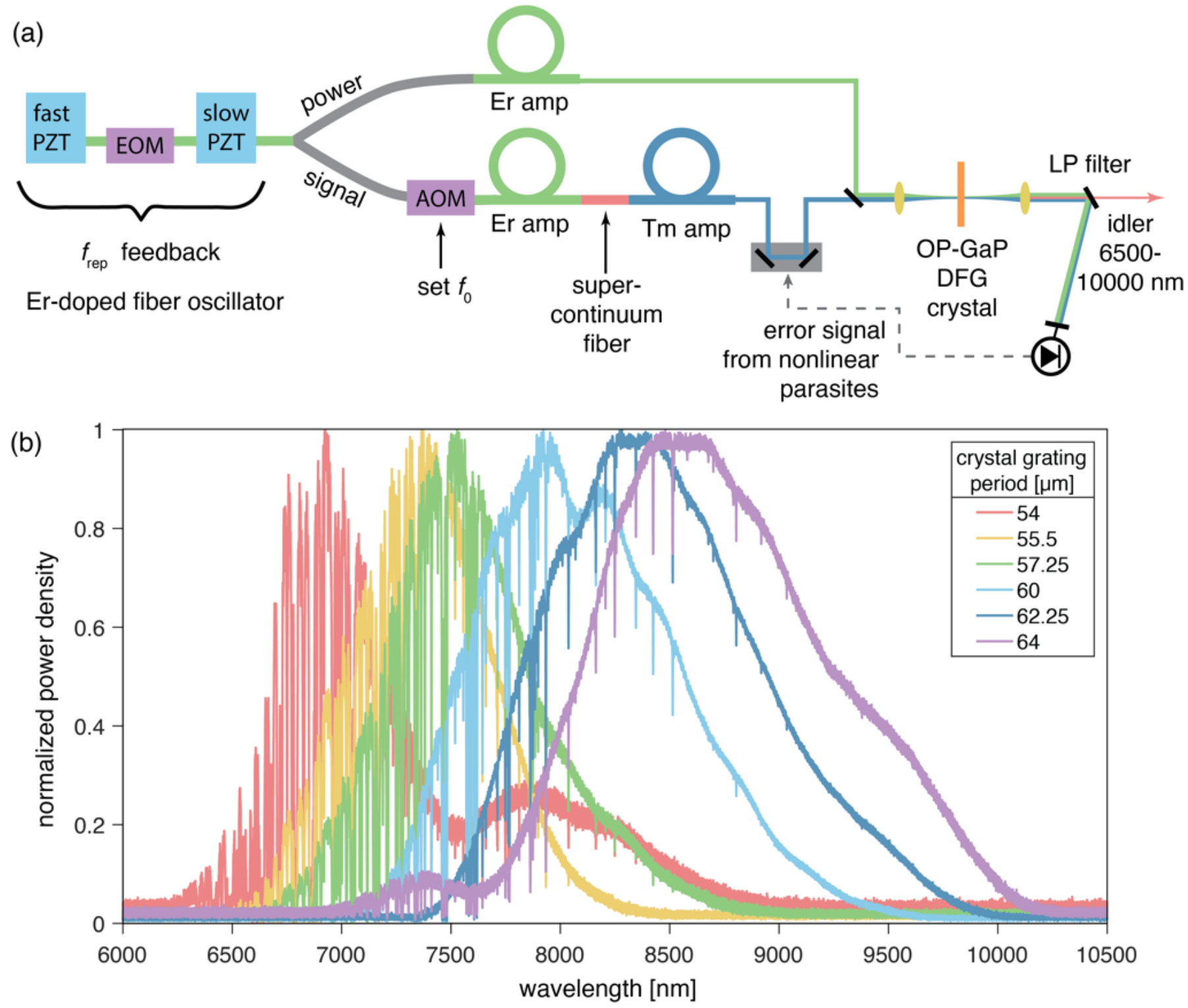

**Fig. 1. (a)** Difference frequency generation (DFG) comb setup used to generate LWIR light between 6500−10000 nm from mixing of the power and signal arms in an orientationally-patterned gallium phosphide (OP-GaP) crystal. PZT = piezoelectric transducer, EOM = electro-optical modulator, AOM = acousto-optical modulator, LP = long-pass. **(b)** Fourier transform spectra of the DFG comb output acquired at different positions of the OP-GaP crystal with normalized spectral power densities. The spectra become noisier at wavelengths shorter than 8000 nm due to absorption by atmospheric water. The grating period of 57.25 µm (green) is used in this study.

The power and signal arms are collimated in free space and combined on a dichroic mirror, after which they are focused with a 75 mm fused silica lens onto an anti-reflection coated orientation-patterned gallium phosphide (OP-GaP) crystal for DFG. The OP-GaP crystal features grating periods of 52.25, 54, 55.5, 57.25, 60, 62.25 and 64 µm. Translation of the crystal in the plane orthogonal to the incident beams allows for selection of the grating period and thereby tuning of the central DFG output wavelength between 6500 and 10000 nm (Fig. 1b). The maximum output power of this system is ~70 mW centered at 8000 nm using the 60 µm grating period. In this work, we use the 57.25 µm grating period and limit the power arm to 2.7 W to yield ~30 mW of DFG light centered at 7500 nm with ~200 cm$^{-1}$ of instantaneous spectral bandwidth. The DFG light is collimated with a ZnSe lens and sent through a long-pass filter with a 3.4 µm cut-off to remove residual power and signal light, as well as any other frequencies generated in the crystal via parasitic nonlinear processes.

We stabilize the DFG output power by controlling the delay between the power and signal pulses to ensure their temporal overlap in the OP-GaP crystal. To achieve this, we apply feedback on an optical delay stage in the signal beam path. The delay stage consists of a home-built retroreflector mounted on a pair of translation stages, one manual (Newport M-UMR5.16) and one piezoelectric (Physik Instrumente P-621.10L). Red light produced by a parasitic sum frequency generation process in the DFG crystal illuminates a silicon photodiode (Thorlabs PDA36A). The resulting signal is maximized at an optical delay slightly offset from the DFG-optimized position, making it suitable for use as an error signal in a side-of-line lock. A proportional-integral controller (Newport LB1005) processes the signal and feeds back on the piezoelectric delay stage through a multi-channel high voltage amplifier (HVA, IMRA America Inc.), closing the loop.

We control the comb $f_{rep}$ by modulating three actuators integrated into the oscillator cavity: an electro-optical modulator (EOM) with a few-MHz bandwidth, a fast piezoelectric transducer (PZT) with a 50 kHz bandwidth, and a slower long-throw PZT. All three actuators require high voltage driving signals, which we provide using additional channels on the IMRA HVA. No active locking is required to stabilize the comb $f_0$, as it is passively canceled in the DFG process [21,22]. However, it is important to be able to introduce a fixed frequency offset to the comb spectrum to compensate for the dispersion of the enhancement cavity mirrors. We must also be able to tune the comb $f_0$ to optimize cavity-coupling in different spectral regions. We use an acousto-optical modulator (AOM, Brimrose) to shift the frequency of the light in the signal arm. The tunable AOM driving frequency is generated by a direct digital synthesizer (DDS, Analog Devices AD9959), the output of which is band-pass filtered and amplified to 30 dBm in two stages.

### 2.2 Vacuum chamber and cryogenic buffer gas cell

Our CBGC apparatus (Fig. 2) is based on a design previously implemented at JILA [23,24]. The cold cell is $(8\ \text{cm})^3$ in volume and is anchored to the 4.2 K stage of a closed-cycle pulse tube helium cryocooler (Cryomech PT410) within a vacuum chamber. The cell is protected from blackbody radiation by a copper shield anchored to the 40 K first stage of the refrigerator, which features a 40 W cooling capacity. We monitor the cell and shield temperatures with silicon diodes (Lakeshore DT-670C-CU diodes and Lakeshore 218E monitor). We place a layer of indium foil at the contact points between the 40 K stage and the shield and between the 4 K stage and the cell to ensure good thermal contact.

The vacuum chamber surrounding the CBGC is evacuated with a turbopump (Pfeiffer Vacuum HiPace 300) backed by a roughing pump (Pfeiffer HiScroll 12). Once the cryocooler reaches temperatures below 10 K, cryosorbs become the main pumping method to trap helium buffer gas [16,25]. We mount two homemade cryosorb plates with a total area of ~1382 $\text{cm}^2$ to the 4 K cell assembly. The cryosorbs consist of activated coconut charcoal (Spectrum Chemical MFG Corp) glued with epoxy (Stycast 2850FT and catalyst 24LV) onto 0.0625” thick copper plates. The cryosorbs can pump helium buffer gas for several hours and achieve a typical base pressure of $2\times10^{-6}$ Torr.

Target molecules are delivered to the CBGC via an aperture in the front face of the cell surrounded by an annular slit through which we flow helium buffer gas regulated with a flow meter (Alicat MC100SCCM) to a typical flow rate of 17 sccm. Helium buffer gas is delivered to the annular slit via copper tubing anchored to the second stage of the cryocooler to achieve some level of pre-cooling. The annular slit is designed to create a curtain of cold molecules that facilitates immediate collisional cooling of the sample. With the cell cold and helium flowing, we see a typical chamber pressure of $2\times10^{-5}$ Torr, corresponding to a helium number density of $10^{13}$–$10^{14}$ $\text{cm}^{-3}$ in the cell. Target molecules ultimately thermalize to temperatures in the range of 10–20 K during their residence time in the cell; see Section 3 for representative spectroscopic measures of molecular temperatures achieved in the CBGC.

Target molecules may be introduced to the chamber via various means. For samples which begin in the gas phase – as ethane does in this work – we use a flow meter (Alicat MC10SCCM) to deliver 0.5–1 sccm of target molecules to the cell aperture via a ¼” stainless steel tube. For samples which begin as a liquid at room temperature – as ethanol does here – we flow 4–6 sccm of room temperature helium through a bubbler containing the liquid of interest, and again deliver this flow to the cell aperture via a ¼” stainless steel tube. The flow from the bubbler to the cell is regulated manually with a needle valve. The flexible design of the setup allows for the introduction of initially gaseous and liquid samples alike, requiring only minor adjustments to switch between these modalities. We can also use a resistively heated copper oven to introduce solid samples into the cell [17,23,24], which we will detail in a future report.

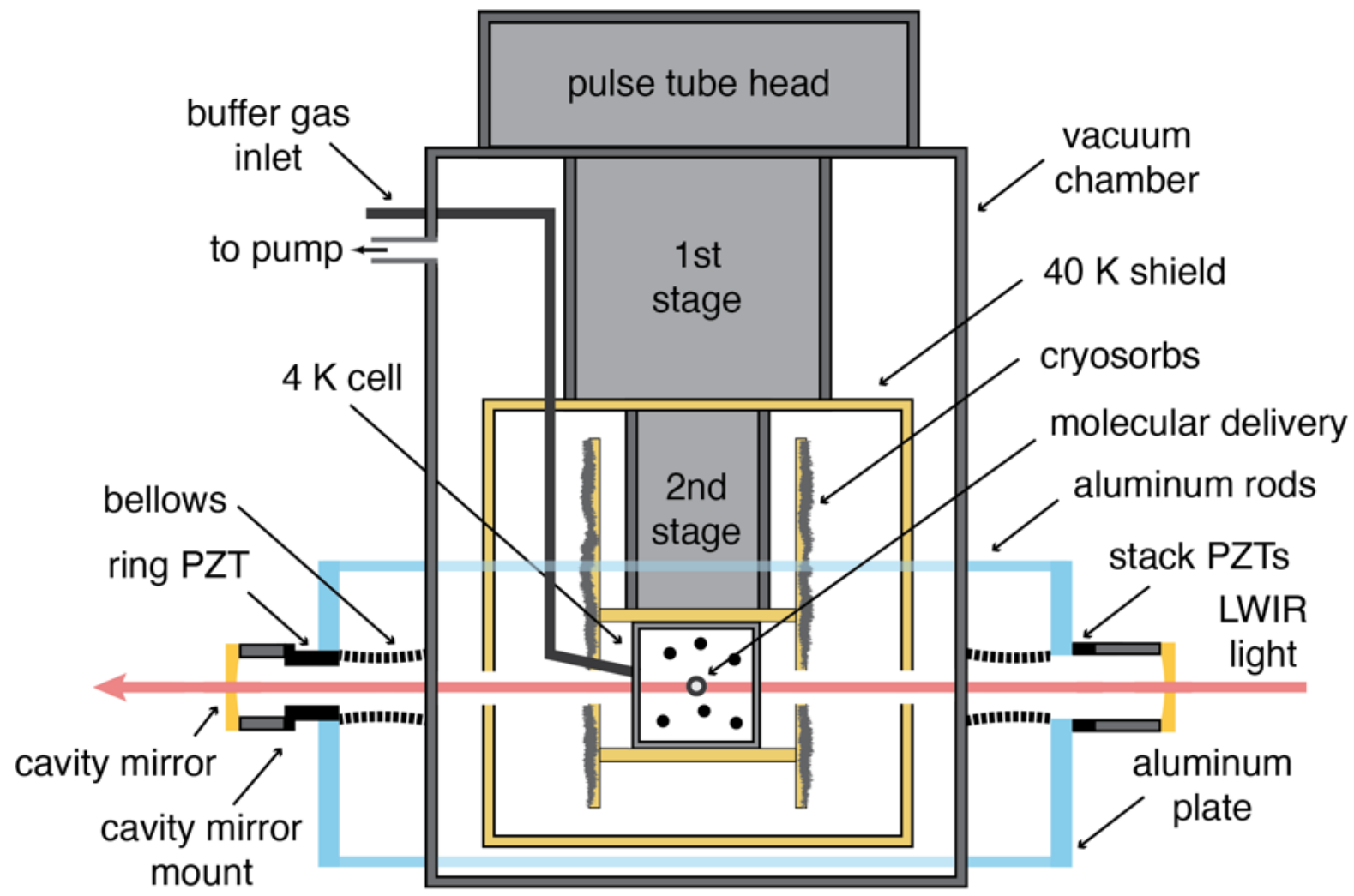

**Fig. 2.** Schematic cross section of the vacuum chamber and cryogenic buffer gas cell setup. PZT = piezoelectric transducer.

## 2.3 Optical cavity and comb coupling

We employ a Fabry-Pérot optical enhancement cavity to increase the interaction length between the comb light and the molecular sample. The optical cavity is integrated into the vacuum chamber that houses the cryogenic buffer gas cell (Fig. 2). The cavity is constructed from two high-reflectivity plano-concave dielectric mirrors on ZnSe substrates with a radius of curvature of 6 m (LohnStar Optics). The mirrors used here feature a coating with a design transmission of 110 ppm and estimated absorption and scattering losses of 50 ppm at a stopband center of 6980 nm. We measure the cavity finesse by analyzing transmitted cavity ringdown traces following the methodology of Poirson *et al.* [26]; at 7000 nm we find $F \approx 14000$ which corresponds to a mirror reflectivity of $R \sim 99.978\%$.

For the data reported here, we use a notch filter (Electro Optical Components Inc.) to select the $7180 \pm 160$ nm region of the LWIR spectrum. At 7180 nm, we measure a reduced cavity finesse of $F \approx 7600$, corresponding to a mirror reflectivity of $R \sim 99.96\%$. In this spectral region, the design specification for mirror transmission is $T = 180$ ppm, yielding an estimate for per-pass mirror losses of $\ell = 220$ ppm. We therefore estimate the circulating intracavity power enhancement factor to be $T/(1-R)^2 = 1125$ in the absence of intracavity absorption. The mirror losses limit the fraction of incident light transmitted through the cavity to $T^2/(1-R)^2 \sim 20\%$. Despite the reduced mirror performance near 7180 nm, this spectral window is of particular interest for the present work because it contains fundamental vibrational bands of ethane and ethanol with a manageable level of water absorption. The notch filter prevents the incoming comb light from exceeding the bandwidth of the mirror stopband where the reflectivity of the coatings fluctuates, leading to unwanted non-resonant transmission, increased noise, and reduced available dynamic range of photodetectors.

The cavity mirrors are mounted in homebuilt 3-point mounts with precision screws (Thorlabs FAS300) allowing fine alignment and manual adjustment of the cavity length. One of the cavity mirror mounts is glued to a high bandwidth ring PZT (PI-USA P-025.10H) for remote modulation and fast feedback on the cavity length. We also incorporate three stack PZTs (Thorlabs PK4FA2P1) to one of the 3-point mounts for longer-throw actuation and locking. The cavity mirror mounts are fixed to two vertically-oriented aluminum plates mounted to the laser table on either side of the vacuum chamber. These aluminum plates are connected via four 1" diameter aluminum rods to stabilize the cavity length. The vacuum chamber itself sits on a separate framework from the laser table to decouple any vibrational noise from the pumps and cryostat from the laser table. A set of bellows vibrationally decouples the cavity mirror mounts from the vacuum chamber.

Efficient coupling of comb light into an optical cavity requires both spatial mode matching and careful frequency stabilization. We mode-match the LWIR beam to the TEM00 cavity mode using a 200 mm focal length Ge lens and a 250 mm focal length ZnSe lens to achieve a beam with a 3.2 mm 1/*e* diameter waist at the center of the cavity. As the LWIR beam is produced with DFG, it has an imperfect beam quality; we measure an $M^2$ value of 2.31 for the beam as it is coupled into the cavity. Nonetheless, we are able to mode-match into the TEM00 mode with an intensity ratio of ~200:1 compared to the next most-intense higher-order transverse mode. We keep the cavity length near 53.4 cm to fulfill the condition $FSR = 3 \times f_{\mathrm{rep}}$, yielding 3:1 Vernier filtering [27] of the comb light to an effective repetition rate $f_{\mathrm{fil}}$ near 280 MHz. The simultaneous bandwidth of light transmitted through the cavity is therefore limited by both the notch filter and the mirror dispersion, requiring fine tuning of both $f_{\mathrm{rep}}$ and $f_0$ to achieve optimal comb-cavity coupling. With the current cavity mirrors, we achieve the broadest instantaneous transmission spectrum by setting the comb $f_0 = 58.2$ MHz using the AOM in the signal arm.

We employ a tight cavity-comb locking scheme, illustrated in Fig. 3, to achieve continuous transmission of comb light through the enhancement cavity for compatibility with FTS readout. We use a Pound-Drever-Hall (PDH) lock to stabilize the comb $f_{\mathrm{rep}}$ so it follows the cavity *FSR* [28]. A second loop, referred to hereafter as the "$f_{\mathrm{rep}}$ lock," stabilizes the 16$^{\mathrm{th}}$ harmonic of $f_{\mathrm{rep}}$ to an absolute radio frequency reference by feeding back on the optical cavity length. All frequency synthesizers used for both locks are referenced to the 10 MHz output signal of a Rb frequency standard (Stanford Research Systems FS725).

For the PDH locking scheme, we first apply a 2.7 MHz modulation produced by a DDS (Analog Devices AD9959) to the comb oscillator's intracavity EOM. This frequency lies outside of the HVA bandwidth, so we amplify it with a resonant home-built LC circuit to achieve sufficient modulation depth. A $BaF_2$ wedged window picks off ~4% of the LWIR comb light backreflected from the input cavity mirror, which is then focused onto a HgCdTe photodiode (Kolmar Technologies KLD-0.5-J1/11/DC/PS). We isolate the modulation frequency with a combination of low- and high-pass filters (Mini Circuits) and amplify the signal (Mini Circuits ZFL-500-BNC) before demodulation with a phase detector (Mini Circuits ZRPD-1+). The resulting error signal is processed with a fast servo controller (Toptica Photonics FALC110), then applied to the fast and slow PZTs in the comb oscillator via the IMRA HVA. The PDH feedback is also applied to the comb oscillator's EOM after processing with a home-built proportional gain circuit; a home-built fast summing amplifier is used to combine this feedback with the original 2.7 MHz PDH modulation before it is applied to the EOM through the IMRA HVA.

The secondary $f_{\mathrm{rep}}$ lock scheme provides absolute stabilization of the comb teeth for precision spectroscopy. To generate the error signal for this lock, we mix the 16$^{\mathrm{th}}$ harmonic of the comb $f_{\mathrm{rep}}$ detected via picked-off oscillator light with a sinusoidal signal near 1.47 GHz generated with a frequency synthesizer (RIGOL DSG815). This mixing process yields a difference signal near 28 MHz which we compare to the output of a second DDS using a phase detector (Analog

Devices AD8302) to generate a phase error signal. This error signal is then fed to a servo controller (Newport LB1005) and into the master input of a PZT controller (Thorlabs MDT693B) that drives both the cavity ring and stack PZTs.

To sample molecular absorption spectra with frequency steps smaller than the cavity-filtered comb spacing of ~280 MHz, we interleave spectra acquired at systematically scanned values of $f_{rep}$. We step $f_{rep}$ by adjusting the DDS output near 28 MHz used to generate the $f_{rep}$ lock error signal, allowing for arbitrarily dense sampling. In the measurements presented here, we interleave fifty spectra acquired by stepping $f_{rep}$ in 12.5 Hz steps to achieve a sample spacing of approximately 5 MHz at 7200 nm. We describe this procedure in more detail in Section 2.4 below.

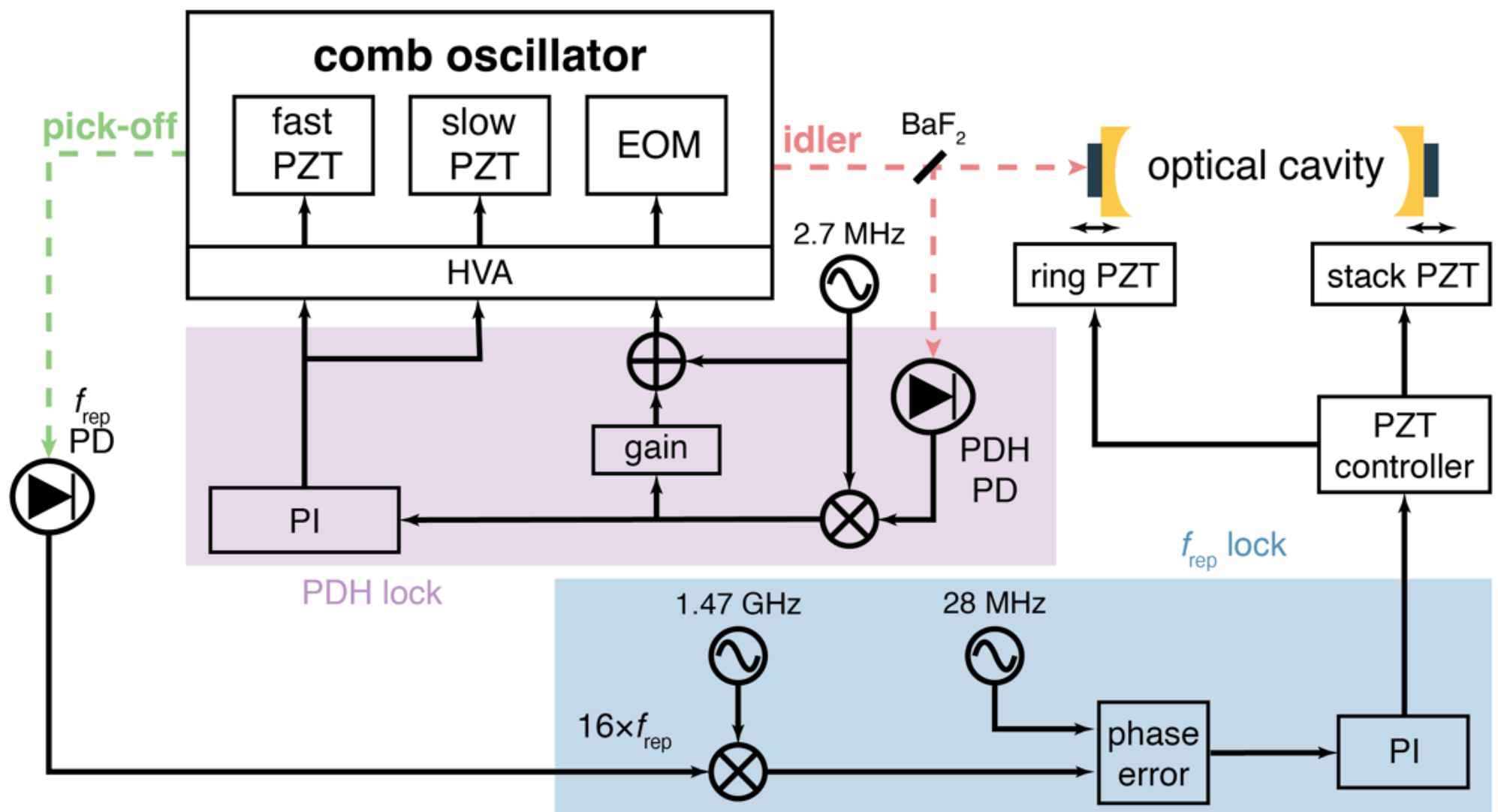

**Fig. 3.** Scheme for cavity-comb stabilization, including both Pound-Drever-Hall (PDH) and $f_{rep}$ locking. PZT = piezoelectric transducer, PD = photodiode, EOM = electro-optic modulator, HVA = high-voltage amplifier, PI = proportional-integral.

### 2.4 Comb-mode resolved Fourier transform spectroscopy

We read out broadband comb spectra with a home-built FTS based on the Michelson interferometer design [29]. We recollimate the beam transmitted through the cavity with a telescope consisting of two Ge lenses with focal lengths of 75 and 50 mm. Comb light then enters the FTS and is directed into the two arms of the interferometer with a 50:50 $CaF_2$ beam splitter coated for 2000-8000 nm (Thorlabs BSW520). Both interferometer arms end in corner cube retroreflectors (Newport UBBR2.5-1S) mounted on either side of the scanning cart of a 1000 mm-long direct drive delay stage (Zaber X-LDQ1000C-AE53D12). This design yields a maximum optical path difference of 4 m and therefore a nominal spectral resolution of 75 MHz. We record LWIR interferograms with a pair of response-matched HgCdTe photodiodes (Kolmar Technologies KLD-0.5-J1/11/DC/PS in both the balanced and unbalanced interferometer outputs. A custom-designed autobalancing circuit (JILA Electronics Lab, Boulder, CO) subtracts the two out-of-phase LWIR interferograms, effectively canceling the common mode intensity noise and doubling the signal contrast.

A 1550 nm cw laser with a <2 kHz instantaneous linewidth (RIO Orion 3375-3-02-4) propagates parallel to the comb beams in the interferometer to serve as an optical frequency reference. A single InGaAs photodiode (Thorlabs PDA10CS2) in the unbalanced interferometer output records the reference interferogram.

We use an oscilloscope card (National Instruments PCI-5922) to digitize the interferograms with 16-bit resolution and a 5 MS/s sampling rate. The FTS delay stage triggers the data acquisition by sending a digital signal to the digitizer, after which the LWIR interferogram is resampled at the zero crossings of the cw laser interferogram.

Traditional FTS instruments are constrained by spectral resolution limited by the maximum optical path difference of the interferometer, $D$. This limit manifests in a frequency-domain instrumental lineshape (ILS) of a sinc function with a width given by:

$$f_{\mathrm{FTS}} = \frac{c}{D} \tag{1}$$

where $c$ is the speed of light. $f_{\mathrm{FTS}}$ also describes the frequency sampling step size of a spectrum produced from digital fast Fourier transform (FFT) of a discretely sampled interferogram. For our acquisition scheme,

$$D = N\,\lambda_{\mathrm{ref}}/2 \tag{2}$$

where $N$ is the number of sampling points in the interferogram and $\lambda_{\mathrm{ref}}$ is the wavelength of the cw reference laser.

The spectrum of an optical frequency comb consists of a series of narrow, evenly spaced lines, which can be treated as Dirac delta functions for the purposes of spectral processing. In FTS spectra, the comb teeth are convolved with the sinc ILS, introducing distortion through both broadening and cross-talk between neighboring comb teeth mediated by the sinc function sidebands. To alleviate this, we employ the sub-nominal resolution scheme first described by Masłowski *et al*. [30,31] to acquire comb-mode resolved spectra free of this ILS limit. This method (a) sets the spacing of the nodes in the sinc ILS to be commensurate with the Vernier-filtered comb repetition rate to minimize cross-talk between comb teeth and (b) optimizes the sampling of the frequency-domain spectrum to extract undistorted comb teeth intensities.

First, we resize the comb interferogram to match the nominal resolution of the digital FFT, $f_{\mathrm{FTS}}$, to the Vernier-filtered comb repetition rate, $f_{\mathrm{fil}}$, by trimming the number of sampled points down to:

$$N = \mathrm{round}\left[\frac{2\,j\,c}{f_{\mathrm{fil}}\ \lambda_{\mathrm{ref}}}\right] \tag{3}$$

where $j$ is a positive integer. After trimming, $j \times f_{\mathrm{FTS}} \approx f_{\mathrm{fil}}$. The resulting trimmed interferogram will contain $j$ interference bursts. The central burst arises from the interference of a laser pulse with itself. For $j$ values larger than 1, $D$ exceeds $c/f_{\mathrm{fil}}$, the optical delay between pulses, giving rise to additional interference events between a given laser pulse and preceding or trailing pulses.

The discrete nature of experimental data leads to a slight (Hz-level) mismatch between $f_{\mathrm{fil}}$ and $f_{\mathrm{FTS}}$. In effect, this mismatch causes the sinc-broadened comb teeth to be sampled on the slopes of their ILS rather than at their maxima, leading to errors in spectral intensity that scale linearly with the comb tooth number $m$. To correct for this, we first calculate $f_{\mathrm{shift}}$, the frequency offset between the discrete FFT sampling points and comb tooth frequencies:

$$f_{\mathrm{shift}} = -m_{\mathrm{opt}} \times \left[f_{\mathrm{FTS}} - \frac{f_{\mathrm{fil}}}{j}\right] \tag{4}$$

where $m_{\mathrm{opt}}$ is the index of a representative comb tooth chosen to lie at the center of the cavity transmission spectrum. Here, $m_{\mathrm{opt}}$ is on the order of $10^5$.

We shift the FFT sampling points by $f_{\mathrm{shift}}$ by multiplying the experimental interferogram by a complex exponential function. This procedure accounts for the comb $f_0$ as well:

$$I_{\text{shift}}(d) = I_0(d)\ \times \exp\left[-\frac{2\pi i(f_0 + f_{\text{shift}})d}{c}\right] \tag{5}$$

where $I_0(d)$ and $I_{\text{shift}}(d)$ represent the original and frequency-shifted interferograms, respectively, and $d$ is the variable optical path difference over which the interferogram is collected.

The accuracy of this sampling correction depends on the reference laser wavelength $\lambda_{\text{ref}}$ per Eqs. (2) and (3). Here, we use an optical spectrum analyzer (Yokogawa AQ6374) to measure $\lambda_{\text{ref}}$ = 1550.07 ± 0.05 nm. Unfortunately, this accuracy is insufficient to ensure correct determination of $N$ and $f_{\text{shift}}$. We solve this in processing by determining the value of $\lambda_{\text{ref}}$ that yields an optimally sampled comb spectrum. We systematically resize and frequency shift the LWIR interferogram, stepping $\lambda_{\text{ref}}$ by ±0.005 nm in each step and recording the subsequent intensity of the comb tooth with index $m_{\text{opt}}$. This procedure traces out the frequency-domain ILS; fitting the central peak of this ILS gives the optimized value of $\lambda_{\text{ref}}$. While the optimization is performed for one specific comb tooth, the mismatch between $f_{\text{fil}}$ and $f_{\text{FTS}}$ is small enough to ensure that the sampling remains close to optimal within the entire spectral bandwidth of our measurements.

This loop to optimize $\lambda_{\text{ref}}$ is unfortunately quite computationally expensive. Careful analysis shows that the optimized $\lambda_{\text{ref}}$ value varies with a standard deviation of 40 ppb over 100 interferograms. This translates to <0.2% error in the determination of the comb tooth intensity, well below other dominant noise sources. We therefore run the $\lambda_{\text{ref}}$ optimization loop only once while processing a full $f_{\text{rep}}$ scan. We construct the preliminary frequency axis by assigning the comb tooth frequencies based on the optimized value of $\lambda_{\text{ref}}$, the locked value of $f_{\text{rep}}$, and the value of $f_0$ set by the AOM; the frequency calibration is later confirmed against the absolute positions of well-characterized water lines (*vide infra*). This process minimizes the effect of crosstalk between the sinc lineshapes of neighboring comb teeth, achieving spectral resolution limited only by the comb tooth linewidth [30]. The optimization of $\lambda_{\text{ref}}$ should not ultimately contribute frequency uncertainty to molecular line positions in our CE-FCS measurements, as we operate in the regime where the molecular linewidths are much narrower than the comb $f_{\text{rep}}$ [31].

With the comb-mode resolved, ILS-free cavity-enhanced transmission spectrum in hand, we now turn towards constructing an absorption spectrum. The experimental comb transmission spectra often contain baseline oscillations from etaloning in the beam path. To filter out these oscillations, we FFT the frequency-domain spectrum, identify sharp peaks corresponding to etalons, and replace them with zeros. We then perform an inverse FFT to return a transmission spectrum with minimized etalon influence, which we refer to hereafter as $T(\nu)$. Next, we construct a reference transmission spectrum, $T_0(\nu)$ via moving average smoothing of $T(\nu)$. The window size of this moving average is selected to minimize distortion of narrow absorption features of cold molecules, but we note that this process can distort the shapes of strong, broad, or congested lines, and can therefore complicate absorption measurements of pressure-broadened atmospheric water lines and room temperature intracavity gases in our experiments. We then calculate absorbance $A(\nu)$ using the Taylor series approximation of the Beer-Bouguer-Lambert law

$$A(\nu) = 1 - \frac{T(\nu)}{T_0(\nu)} \tag{6}$$

which is valid for the small absorption signals measured in our cold spectra.

We perform averaging in two steps to account for slow drifts in the spectral envelope of the transmitted comb spectrum. We first average the transmission spectra $T(\nu)$ taken in rapid succession at a single value of $f_{rep}$, since the transmitted comb envelope does not express large variations in shape on the order of seconds to minutes. We perform this averaging after optimizing $\lambda_{ref}$. We average data acquired across separate $f_{rep}$ scans by first calculating $A(\nu)$ according to Eq. (6) to remove the spectral envelopes which do drift from scan to scan on the order of tens of minutes to hours. Finally, we interleave the $A(\nu)$ spectral traces acquired for each value of $f_{rep}$ to arrive at a final, fully-sampled absorption spectrum.

We confirm the absolute calibration of the frequency axis for the comb absorption spectra with reference to the positions of known water lines from the HITRAN database [32–34]. Strong absorption lines appear in the comb spectra from room-temperature, pressure-broadened water vapor in the beamline outside the vacuum chamber. We identify a set of non-saturated transitions in the experimental $T(\nu)$ transmission spectra which feature good signal-to-noise to use as calibrants and use a third-degree polynomial to remove slowly varying baselines, then fit each line with a Voigt profile to find the center frequency. We fit the lineshapes separately for spectra acquired at each $f_{rep}$ value, averaging the final line positions to increase confidence. A linear regression of the fitted experimental line positions to the HITRAN data yields a scaling factor for the frequency axis of the comb spectrum. The resulting frequency axis allows for the determination of the closest comb tooth to each spectral point and a final frequency assignment. This calibration procedure confirms the comb tooth numbers pre-assigned at the sinc sampling optimization step for all measurements reported here, with the scaling factor yielding frequency points shifted from their pre-assigned positions by no more than 15% of the comb $f_{rep}$.

We estimate the noise-equivalent absorption (NEA) per spectral element for our measurements according to [35,36]:

$$\alpha_{\min} = \frac{\sigma}{L_{\text{eff}}}\sqrt{\frac{t}{M}} \tag{7}$$

where $\sigma$ is the standard deviation of a small baseline region of the absorption spectrum containing no molecular lines, and with etalons filtered out; $L_{\text{eff}} = 2F/\pi \times L$ is the effective cavity-enhanced absorption path length, where we take the single-pass intracavity absorption pathlength $L$ to be that of the cold cell (8 cm); $t$ is the total data acquisition time; and $M$ is the number of spectral elements in the interleaved spectrum. Here, finding $\sigma \sim 0.001$ for data collected over $M \sim 270{,}000$ spectral points for $t = 2{,}250$ seconds, we obtain an NEA per spectral element of $\alpha_{\min} = 2.5 \times 10^{-9}$ cm$^{-1}$ Hz$^{-1/2}$, comparable performance to other mid-IR CE-FCS demonstrations [23,37]. Finally, we note that atmospheric water vapor presents a major challenge for spectroscopy near 7200 nm. We make use of an enclosure around the entire laser beamline and FTS system which can be purged with dry nitrogen to minimize water absorption lines in the resulting comb spectra.

## 3. Results and Discussion

We now present LWIR CE-FCS measurements of ethane and ethanol acquired both at room temperature and under cryogenic conditions. Our central focus is on ethane, for which we retrieve rovibrational line positions and linewidths of transitions in the $\nu_6$ fundamental band and benchmark against available literature data. We use the cold ethane spectra to characterize the rotational and translational temperatures achievable in the CBGC. We also present data on ethanol as a more complex system featuring congested rovibrational structure not previously resolved in this spectral region.

### 3.1 Ethane

We target ethane as a first test system for the LWIR CE-FCS apparatus due to its strong absorption near 7200 nm, the availability of reference data, and its importance in both atmospheric chemistry and astrochemistry. Ethane is a prolate symmetric top with $D_{3d}$ symmetry. Here we focus on its $A_{2u}$-symmetric $\nu_6$ rovibrational band lying near 1380 cm$^{-1}$, which corresponds to $CH_3$ deformation. This band features rovibrational selection rules of $\Delta J = 0, \pm 1$ and $\Delta K = 0$ where $J$ is the total rotational angular momentum quantum number, and $K$ is the projection of $J$ on the molecular $C_3$ axis [38].

In Fig. 4, we present CE-FCS absorption spectra of the ethane $\nu_6$ band at both room temperature and with the CBGC cooled to 5 K. Grayed-out spectral regions correspond to absorption by atmospheric water in the LWIR beam path. All spectra are constructed by interleaving data taken at 50 stepped values of the comb $f_{\text{rep}}$, yielding ~5 MHz spectral sampling of the spectrum.

The room temperature ethane spectra in Fig. 4ab feature strong absorption lines as the entire path length of the 53.4 cm-long enhancement cavity fills with gas when the CBGC is not cooled. We estimate an intracavity ethane number density of ~$5\times10^{15}$ cm$^{-3}$ based on the chamber pressure reached under these conditions. In combination, the long pathlength and high number density lead to high signal:noise in the resulting absorption spectra achieved with only 3 interferograms acquired at each value of $f_{\text{rep}}$. However, the spectral congestion evident at room temperature (Fig. 4b) makes line fitting difficult under these conditions.

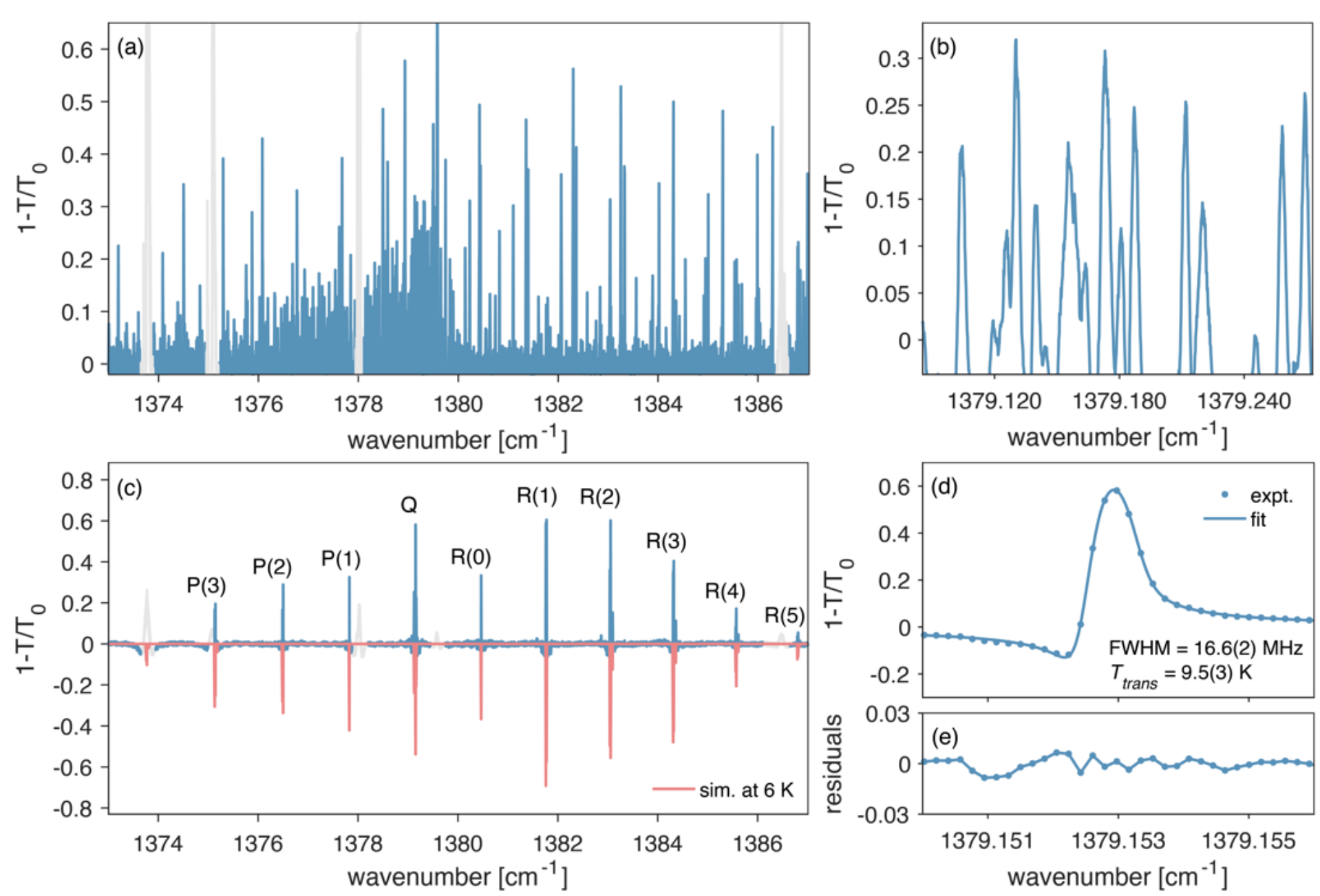

**Fig 4.** CE-FCS spectra of **(a,b)** room temperature and **(c)** cryogenically-cooled ethane. Experimental data are plotted in blue, while grayed out regions correspond to absorption from atmospheric water vapor in the beamline. In panel (c), the inverted pink trace is a cavity-enhanced absorption spectrum of ethane simulated at 6 K by processing the linelist available from HITRAN using the PGOPHER software package and Eq. (8). **(d)** Dispersive lineshape fit to the $J'' = 1 \rightarrow J' = 1$, $K = 1$ Q-branch transition of cold ethane at 1379.152861(3) cm$^{-1}$. **(e)** Residuals for the dispersive fit in panel (d).

By contrast, the spectrum of cryogenically-cooled ethane presented in Fig. 4c is drastically simplified as population is condensed into the lowest-lying rovibrational states. For these measurements, we acquire the data over several scans, resulting in a total of 15 interferograms averaged per $f_{rep}$ step and a total acquisition time of 40 minutes.

We compare our measured spectrum of cold ethane directly against reference data from the HITRAN database [34]. We process the HITRAN linelist for ethane with the PGOPHER software package [39] to calculate the absorption cross section, $\sigma_{\mathrm{HITRAN}}(\nu)$, at 6 K and convolute it with the appropriate Doppler-broadened lineshape. For improved comparison with experiment, we account for cavity-enhancement of absorption signals, making the simplifying assumption that all comb teeth are coupled perfectly on resonance and neglecting any molecular dispersion [40]:

$$A_{\mathrm{HITRAN}}(\nu) = 1 - \frac{T^2 e^{-\sigma_{\mathrm{HITRAN}}(\nu)NL}}{(1 - Re^{-\sigma_{\mathrm{HITRAN}}(\nu)NL})^2} \tag{8}$$

where $T$ and $R$ are the transmission and reflectivity probabilities for a single cavity mirror; $N$ is the intracavity number density; and $L$ is the single-pass intracavity absorption pathlength. The resulting simulation is plotted upside-down in pink in Fig. 4c, and yields excellent agreement with experimental results.

We assign transitions by comparison with HITRAN and label them in clusters according to the corresponding branch (P, Q, or R) and lower rotational quantum number $J''$. We observe a transition at 1380.465195(4) cm$^{-1}$, which we assign as the $J'' = 0 \rightarrow J' = 1$, $K = 0$ transition (labeled R(0) in Fig. 4c). HITRAN reports this line at 1380.46520 cm$^{-1}$ but appears to be missing the associated statistical weights. We assign the statistical weights for this line following the work of Šimečková *et al.* [41] before simulating the final spectrum shown in pink in Fig. 4c. We estimate a number density of $N \sim 2\times10^{12}$ cm$^{-3}$ of cold intracavity ethane based on comparison of the observed cavity-enhanced absorption signals with HITRAN reference data, taking the intracavity absorption path length $L$ to be that of the cold cell (8 cm). We note that uncertainty in the exact intracavity absorption pathlength represents a limiting factor in how well we can constrain the intracavity number density.

To report accurate spectral positions for the rovibrational lines of cold ethane, we must account for the dispersive molecular lineshapes that are well-known to arise in cavity-enhanced comb spectroscopy. Fig. 4d zooms in on the $J'' = 1 \rightarrow J' = 1$, $K = 1$ Q-branch transition of ethane at 1379.152861(3) cm$^{-1}$ as a representative example. This transition's asymmetric lineshape arises from the dispersion of intracavity molecules shifting the positions of the cavity modes, thereby leading to imperfect comb-cavity coupling across the comb spectrum. We account for this by fitting each line of interest with a dispersion-adjusted Gaussian lineshape using the model described by Foltynowicz *et al.* [42]. We disregard pressure broadening as the low chamber pressures used during cryogenic experiments are expected to contribute a negligible Lorentzian linewidth of <100 Hz. We achieve 1σ statistical fit uncertainties for ethane ranging from 100 kHz to 2 MHz in line position and from 200 kHz to 2 MHz in linewidth, depending on the signal to noise ratio of each transition.

**Table 1.** Line positions, full-width-at-half-maximum (FWHM) linewidths, and assignments for rovibrational transitions in the $\nu_6$ fundamental band of ethane measured in this work. The error values provided in our fitted line positions and linewidths represent 1σ statistical fit uncertainties. Our findings are compared against literature data from di Lauro *et al*. made available via HITRAN [43], whose stated line position uncertainties are 0.0001−0.001 $cm^{-1}$ (3−30 MHz). The last column represents the difference between line positions reported in this work and HITRAN line positions.

| | $J''$ | $J'$ | $K$ | Line position, this work [$cm^{-1}$] | FWHM, this work [MHz] | Line position, HITRAN [$cm^{-1}$] | Difference in line positions [MHz] |
|---|---|---|---|---|---|---|---|
| P(3) | 3 | 2 | 0 | 1375.131275(4) | 16.4(3) | 1375.13107 | 6.2 |
| | 3 | 2 | 1 | 1375.144918(9) | 16.1(4) | 1375.14484 | 2.4 |
| P(2) | 2 | 1 | 0 | 1376.487039(4) | 15.8(3) | 1376.48706 | −0.6 |
| | 2 | 1 | 1 | 1376.500789(8) | 17.7(4) | 1376.50081 | −0.6 |
| P(1) | 1 | 0 | 0 | 1377.828040(4) | 16.7(2) | 1377.82675 | 38.7 |
| Q | 2 | 2 | 1 | 1379.12308(6) | 17(3) | 1379.12288 | 6.1 |
| | 1 | 1 | 1 | 1379.152861(3) | 16.6(2) | 1379.15289 | −0.9 |
| | 2 | 2 | 2 | 1379.163975(9) | 16.7(4) | 1379.16386 | 3.5 |
| R(0) | 0 | 1 | 0 | 1380.465195(4) | 16.6(2) | 1380.46520 | −0.2 |
| R(1) | 1 | 2 | 0 | 1381.761394(3) | 17.8(2) | 1381.76122 | 5.2 |
| | 1 | 2 | 1 | 1381.775012(3) | 17.7(2) | 1381.77496 | 1.6 |
| R(2) | 2 | 3 | 0 | 1383.042608(3) | 17.1(2) | 1383.04226 | 10.4 |
| | 2 | 3 | 1 | 1383.056123(3) | 16.9(3) | 1383.05628 | −4.7 |
| | 2 | 3 | 2 | 1383.09674(4) | 17(3) | 1383.09665 | 2.8 |
| R(3) | 3 | 4 | 0 | 1384.308942(3) | 17.7(2) | 1384.30884 | 3.0 |
| | 3 | 4 | 1 | 1384.322270(5) | 17.7(3) | 1384.32219 | 2.4 |
| | 3 | 4 | 2 | 1384.36225(5) | 18(2) | 1384.36234 | −2.7 |
| R(4) | 4 | 5 | 0 | 1385.560292(9) | 18.6(6) | 1385.56031 | −0.6 |
| | 4 | 5 | 1 | 1385.573460(8) | 18.6(4) | 1385.57339 | 2.1 |
| R(5) | 5 | 6 | 0 | 1386.80966(2) | 21(2) | 1386.80973 | −2.0 |

Table 1 summarizes the fitted line positions for all ethane transitions considered in this work and compares them to the best available literature data from HITRAN. The HITRAN data is sourced from experimental and computational studies by di Lauro *et al.* and Lattanzi *et al.* [34,38,43] with stated uncertainties of 0.001−0.0001 cm$^{-1}$ (3−30 MHz). Most of the positions we report here agree with literature values within 0.0001−0.0002 cm$^{-1}$ (3−6 MHz). One exception is the $\nu_6$, $J'' = 1 \rightarrow J' = 0$, $K = 0$ P-branch transition at 1377.828040(4) cm$^{-1}$ which lies close to a strong water transition. The baseline distortion caused by water could cause a systematic shift, potentially explaining the larger (38.7 MHz) discrepancy from the position reported in HITRAN.

We now use the cold ethane spectra to characterize the molecular temperatures reached in our CBGC. First, we use the average dispersion-corrected Doppler widths of all lines to calculate an average translational temperature of $T_{\mathrm{trans}} = 10.1 \pm 0.5$ K, which is consistent with previous spectroscopic studies performed in CGBCs [24]. In addition, we can determine the rotational temperature, $T_{\mathrm{rot}}$, using our measured relative intensities of R-branch transitions, along with the Einstein $A$ coefficients and state energies from HITRAN. In particular, we follow the formalism for the relationship between line intensity and Einstein $A$ coefficients described by Šimečková *et al.* [41]. In principle, the ratio of the peak absorption intensities of two neighboring rovibrational transitions labeled $a$ and $b$ should obey:

$$\frac{I_b}{I_a} = \frac{g_b'}{g_a'} \times \frac{A_b}{A_a} \times e^{-\frac{\Delta E}{k_B T_{\mathrm{rot}}}} \times \frac{\nu_a^3}{\nu_b^3} \tag{9}$$

where $I_{a,b}$ is the experimentally-measured intensity of transition $a$ or $b$ retrieved from our dispersion-adjusted lineshape fitting; $g_{a,b}'$ is the statistical weight of the upper state of each transition; $A_{a,b}$ is the Einstein $A$ coefficient for each transition; $\Delta E = E_b'' - E_a''$ is the difference in energy between the lower states of the two transitions; $k_{\mathrm{B}}$ is the Boltzmann constant; and $\nu_{a,b}$ is the frequency of the transition. $T_{\mathrm{rot}}$ can be found by fitting the slope of $\ln\left(\frac{I_b}{I_a} \times \frac{g_a'}{g_b'} \times \frac{A_a}{A_b} \times \frac{\nu_b^3}{\nu_a^3}\right)$ vs. $\Delta E$. Applying this method to the 12 R-branch transitions from R(0) to R(4), we find $T_{\mathrm{rot}} = 5.5 \pm 0.4$ K for ethane. This rotational temperature is consistent with previous work on cryogenic buffer gas cooling of atoms and molecules [23,44]. These results demonstrate successful cooling of both the translational and rotational degrees of freedom of ethane gas in the CBGC.

### 3.2 Ethanol

We next target the absorption bands of ethanol near 7200 nm to further test the capabilities of our apparatus. There is a distinct lack of high-resolution gas phase spectroscopic data for ethanol with mainly low-resolution absorption cross section data available across the literature [45–47] and in the HITRAN database [48,49]. Konnov *et al.* [50] recently reported high-resolution spectra of ethanol with a dual-comb spectrometer in the 845−1310 cm$^{-1}$ range, but did not venture to higher frequencies.

Ethanol appears as two conformers: the *trans* conformer has $C_s$ symmetry while the *gauche* is $C_1$ [51]. We focus here on the $\nu_{10}$ band of the *gauche* conformer centered near 1394 cm$^{-1}$, which corresponds to a $CH_3$ symmetric stretching deformation [52]. CE-FCS absorption spectra of ethanol in this region are presented in Fig. 5 for both a room temperature sample and with the CBGC cooled to 5 K. In both cases, ethanol vapor is delivered to the chamber via bubbler, as described in Section 2.2. The room temperature spectrum of ethanol (Fig. 5a) is much more congested than that of ethane, owing to its larger size and lower symmetry.

The cold ethanol spectra plotted in Fig. 5bc are considerably simpler. We acquire the data over 12 scans, with two $f_{rep}$ averages per scan for a total of 24 averages and 65 minutes of acquisition time. Fitted dispersive Gaussian profiles (Fig. 5d) yield linewidths in the range of 18−19.5 MHz which yield accordingly translational temperatures of $T_{trans}$ = 18−20 K. $T_{trans}$ is larger here than it was in the ethane measurements, which we attribute to the larger molecular mass of ethanol and the flow of warm helium carrier gas into the cell from the bubbler. We achieve statistical uncertainties for line positions on the order of 0.3 MHz for non-overlapping transitions. We provide a linelist for the rovibrational transitions measured here for the $\nu_{10}$ band of *gauche*-ethanol in Table S1 of the Supplementary Material, though assignment and detailed analysis of these transitions are beyond the scope of the current report. The spectra reported for ethanol in the paper represent the first high resolution spectra of this molecule near 7200 nm under either room temperature or cryogenic conditions.

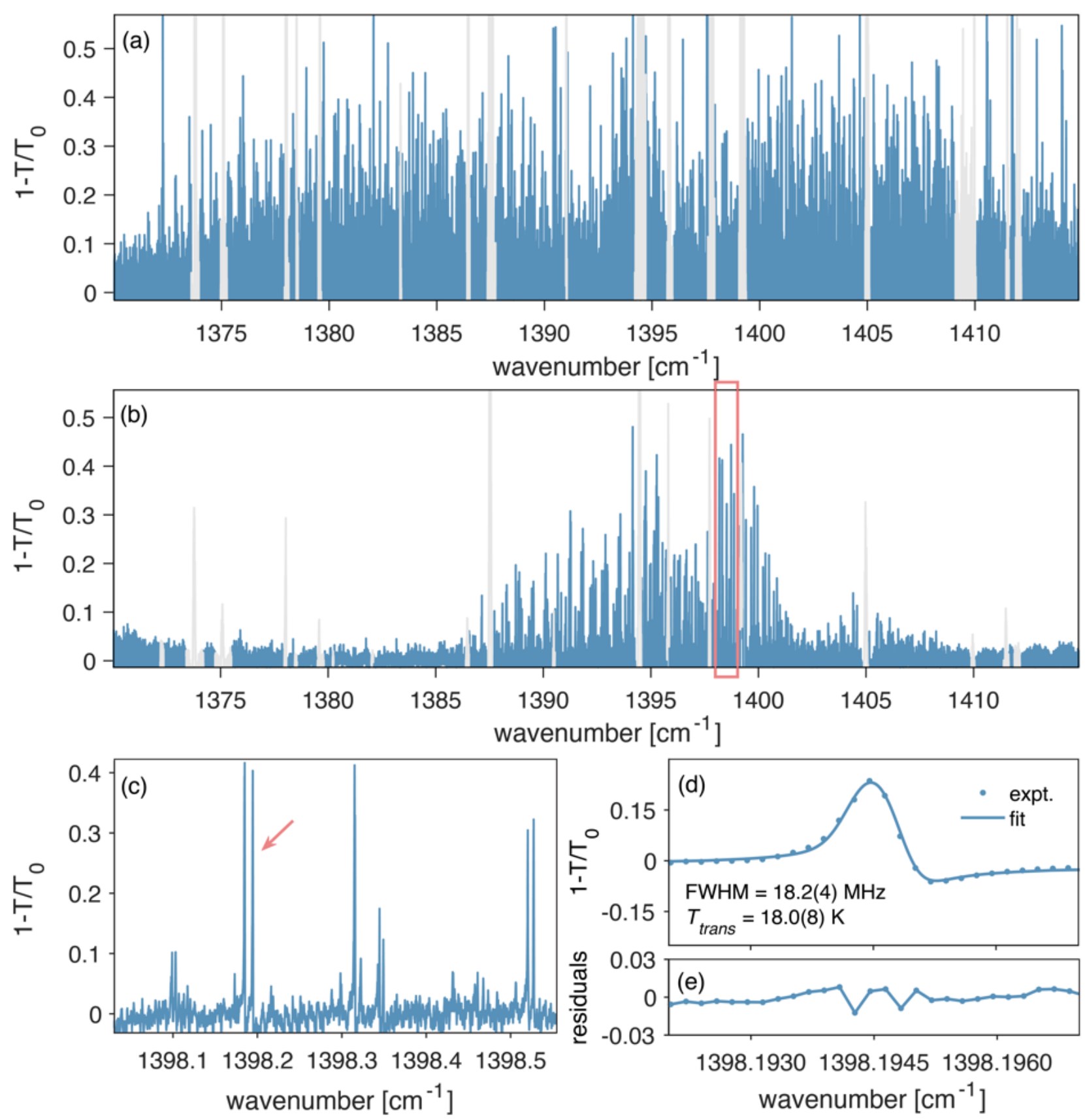

**Fig 5.** CE-FCS overview spectra of **(a)** room temperature and **(b,c)** cryogenically-cooled ethanol, in the region of the the $\nu_{10}$ band of the *gauche* conformer. Panel (c) zooms in on the region in panel (b) boxed within the orange rectangle. Experimental data are plotted in blue, while grayed out regions correspond to absorption from atmospheric water vapor in the beamline. **(d)** Representative dispersive Gaussian lineshape fit to the absorption line of cold ethanol at 1398.19458(3) $cm^{-1}$ indicated with an arrow in panel (c). The line shown here features different absorption intensity from the data shown in panel (c) as is taken from a different dataset acquired without purging the laser enclosure with nitrogen, albeit with more averaging. **(e)** Residuals for the dispersive fit in panel (d).

## 4. Conclusions

We report the development of a new cavity-enhanced frequency comb spectroscopy instrument coupled with cryogenic molecular cooling that will serve as a platform for precision spectroscopy of large hydrocarbons and other complex molecules. Our setup yields broadband, high resolution, and high sensitivity spectra in the LWIR region, ideal for the study of fundamental vibrational transitions. We make use of a sub-nominal resolution technique to process FTS data, achieving spectral precision limited only by the comb tooth linewidth. This allows us to sample spectra densely enough to resolve the narrow, congested absorption lines of complex molecules.

We showcase the capabilities of this setup by presenting broadband absorption spectra of ethane and ethanol near 7200 nm. In ethane, we improve upon existing reference data, reporting rovibrational line positions in the $\nu_6$ band with higher precision than previous demonstrations. Meanwhile, we resolve rovibrational lines for the first time in the $\nu_{10}$ band of *gauche*-ethanol; interpretation and assignment of these lines will be discussed in follow-up work.

The lack of high-resolution laboratory measurements remains a chief limitation in the detection and assignment of new molecules in the interstellar medium and other astrophysical environments. Precision LWIR spectroscopy can provide accurate rotational constants to guide microwave spectroscopy and radioastronomy, which remain the primary methods to detect molecules in space. In future work, we plan to extend this LWIR comb spectrometer to perform broadband, high-resolution absorption spectroscopy of the fundamental vibrational features of larger astrochemically-relevant species, including polycyclic aromatic hydrocarbons and fullerenes.

### Funding

National Science Foundation (AST-2307443).

### Disclosures

The authors declare no conflicts of interest.

### Data Availability

The data that support the findings of this study are available from the corresponding author upon reasonable request.

### Acknowledgments

We thank Glenn Atkinson, Patrick Bradshaw, and Matt Komor of the Princeton University Physics Department Machine Shop for their help in constructing the vacuum chamber assembly, Stanley Chidzik of the Princeton University Department of Physics Electronics Shop for his help with troubleshooting electronics, and Darryl Johnson of the Princeton University Department of Physics for his generous assistance with materials and machining. We also thank Terry Brown and Ivan Ryger of the JILA Electronics Lab for their help in fabricating and troubleshooting the interferometer autobalancing circuit.

### Supplemental Document

See Supplementary Material for Table S1, a rovibrational linelist for cold ethanol.

# SUPPLEMENTARY MATERIAL

**Long-wave mid-infrared cavity-enhanced frequency comb spectroscopy of cold, complex molecules**

Negar Baradaran,[1,†] Dominik Charczun,[1,†] Tanay Nambiar,[1] Meijun Zou,[1]
Kevin F. Lee,[2] Martin E. Fermann,[2] and Marissa L. Weichman[1,*]

[1]Department of Chemistry, Princeton University, Princeton, New Jersey, 08544, USA
[2]IMRA America Inc., 1044 Woodridge Ave, Ann Arbor, Michigan 48105, USA
†These authors contributed equally.
[*]weichman@princeton.edu

**Table S1.** Line positions and full-width-at-half-maximum (FWHM) linewidths, for rovibrational transitions in the $\nu_{10}$ band of *gauche*-ethanol measured in this work. The error values provided in our fitted line positions and linewidths represent 1σ statistical fit uncertainties.

| Fitted Position [cm$^{-1}$] | FWHM [MHz] | Fitted Position [cm$^{-1}$] | FWHM [MHz] |
|---|---|---|---|
| 1388.10287(2) | 18(1) | 1393.4108(1) | 17.0(6) |
| 1388.10742(6) | 21(5) | 1393.42439(4) | 16(3) |
| 1388.27441(3) | 20(2) | 1393.42797(5) | 20(4) |
| 1388.28649(4) | 18(2) | 1393.48064(5) | 21(3) |
| 1388.30032(2) | 18(2) | 1393.50098(3) | 16.6(7) |
| 1388.31263(2) | 18(2) | 1393.56901(2) | 18.7(9) |
| 1388.72037(4) | 18(2) | 1393.574(2) | 20.5(8) |
| 1388.84162(4) | 18(4) | 1393.58363(1) | 17.4(5) |
| 1388.8552(4) | 17(2) | 1393.58723(3) | 20(2) |
| 1388.86492(2) | 19.3(7) | 1393.80612(3) | 18(2) |
| 1388.88475(1) | 18.3(6) | 1393.81355(3) | 17(2) |
| 1388.8984(3) | 21(2) | 1393.94588(4) | 21(3) |
| 1388.93064(6) | 19(3) | 1393.95979(5) | 23(4) |
| 1388.94619(1) | 20.6(7) | 1394.9515(6) | 19(4) |
| 1388.95507(4) | 16(3) | 1394.95489(2) | 22(1) |
| 1388.98669(2) | 24(3) | 1394.99576(5) | 21(2) |
| 1389.00653(3) | 18(2) | 1395.00054(7) | 22(3) |
| 1389.33016(3) | 20(2) | 1395.13391(2) | 19.3(8) |
| 1389.40686(2) | 16(1) | 1395.15026(4) | 19(2) |
| 1389.45751(3) | 18(2) | 1395.18471(1) | 19.5(5) |
| 1389.46736(2) | 19(1) | 1395.25652(2) | 20(2) |
| 1389.4826(2) | 17.9(8) | 1395.27346(2) | 18.5(5) |
| 1389.52542(2) | 20.0(9) | 1395.34757(7) | 21(4) |
| 1389.54215(3) | 22(1) | 1395.35696(1) | 18.7(5) |
| 1389.54913(4) | 23(3) | 1395.53182(3) | 19(1) |
| 1389.55393(1) | 18.3(6) | 1396.08131(7) | 23(5) |
| 1389.56244(2) | 19.9(9) | 1396.09469(1) | 19.9(6) |
| 1389.57212(4) | 19(2) | 1396.24307(2) | 20.1(9) |
| 1390.04828(2) | 19(2) | 1396.26278(3) | 20(1) |
| 1390.06357(6) | 22(2) | 1396.35188(2) | 20(2) |
| 1390.09996(3) | 18(1) | 1396.51731(6) | 17(3) |
| 1390.10966(2) | 19.4(7) | 1396.56932(4) | 22(2) |
| 1390.117(3) | 20(1) | 1396.64182(2) | 20.4(7) |
| 1390.12599(3) | 21(1) | 1396.77137(2) | 21(1) |
| 1390.18225(2) | 20(1) | 1396.78612(3) | 22(2) |
| 1390.19236(3) | 19(2) | 1396.8948(2) | 22(1) |
| 1390.67787(7) | 18(4) | 1396.90586(2) | 19.1(8) |
| 1390.6953(6) | 21(5) | 1397.07641(2) | 18.6(6) |
| 1391.19618(3) | 19(1) | 1397.08583(2) | 21(1) |
| 1391.20362(6) | 16(5) | 1397.19976(1) | 21.5(8) |
| 1391.21132(2) | 19(1) | 1397.20457(3) | 19(2) |
| 1391.21756(2) | 21.2(8) | 1397.2136(2) | 21(1) |
| 1391.25848(5) | 20(2) | 1397.22209(2) | 19(1) |
| 1391.27644(2) | 19(1) | 1397.29759(4) | 19(3) |
| 1391.4922(2) | 19(1) | 1397.30876(5) | 22(3) |
| 1391.49574(4) | 18(2) | 1397.31988(3) | 22(2) |
| 1391.61366(5) | 17(4) | 1398.18501(2) | 18.3(6) |
| 1391.63347(3) | 22(2) | 1398.19458(3) | 18.2(4) |
| 1391.63527(3) | 16(2) | 1398.34928(2) | 17.9(8) |
| 1391.65916(2) | 19(1) | 1398.52039(2) | 18.7(8) |
| 1391.66189(2) | 18(2) | 1398.52733(2) | 20(1) |
| 1391.69358(2) | 20(2) | 1398.72925(2) | 19.2(9) |
| 1391.74927(2) | 18.5(9) | 1398.73889(3) | 19(2) |
| 1391.75473(2) | 22.3(7) | 1399.65983(2) | 18.4(6) |
| 1391.76231(2) | 22.1(6) | 1399.6673(5) | 22(3) |
| 1391.77428(6) | 18(3) | 1399.79999(2) | 18.9(7) |
| 1391.77594(4) | 21(2) | 1399.80992(2) | 18.9(7) |
| 1391.82747(2) | 17.5(5) | 1399.95871(2) | 18(1) |
| 1391.83947(2) | 17.6(8) | 1399.96328(3) | 17(2) |
| 1391.85766(8) | 19(5) | 1400.00453(2) | 19(1) |
| 1391.8898(6) | 16(4) | 1400.21269(3) | 21(1) |
| 1392.13965(2) | 19(1) | 1400.23205(2) | 21(1) |
| 1392.18106(3) | 19(1) | 1400.32563(1) | 16.9(8) |
| 1392.19315(3) | 19(1) | 1400.33781(2) | 16.9(8) |
| 1392.23919(5) | 19(2) | 1400.34126(2) | 21(1) |
| 1392.2546(9) | 19(5) | 1400.49015(3) | 22(1) |
| 1392.259(3) | 18(2) | 1400.49778(6) | 17(2) |
| 1392.26323(4) | 16(2) | 1400.56665(2) | 22(1) |
| 1392.2679(5) | 18(4) | 1400.71686(2) | 21(1) |
| 1392.31429(4) | 22(2) | 1400.85942(2) | 16(1) |
| 1392.32078(2) | 17.3(7) | 1400.91465(4) | 20(3) |
| 1392.33309(6) | 20(3) | 1400.951592(9) | 16.0(6) |
| 1392.35111(1) | 18.0(9) | 1400.98018(3) | 17(2) |
| 1392.40530(2) | 18.0(9) | 1401.00000(4) | 20(2) |
| 1392.41522(2) | 17.5(8) | 1401.08627(1) | 23(1) |
| 1392.41921(2) | 17.3(9) | 1401.27821(2) | 22(2) |
| 1392.4297(2) | 17.7(7) | 1401.36648(8) | 23.6(6) |
| 1392.43766(8) | 16(7) | 1401.39106(2) | 18(1) |
| 1392.72548(2) | 17.9(8) | 1401.4952(1) | 16.7(7) |
| 1392.73744(3) | 20(2) | 1401.59294(1) | 19(1) |
| 1392.77022(4) | 18(2) | 1401.66775(1) | 17.1(8) |
| 1392.81394(4) | 19(5) | 1402.07772(2) | 18(1) |
| 1392.82182(2) | 18(1) | 1402.14517(1) | 16.8(8) |
| 1392.84357(2) | 19.7(9) | 1402.75319(4) | 17(2) |
| 1392.87204(2) | 22(1) | 1403.24365(2) | 20(1) |
| 1392.8825(4) | 19(2) | 1403.82398(2) | 20(2) |
| 1392.88493(4) | 23(3) | 1403.86095(2) | 19(1) |
| 1392.88917(4) | 21(3) | 1403.900275(8) | 16.5(6) |
| 1392.89130(4) | 20(3) | 1403.99688(2) | 18(1) |
| 1392.90141(6) | 16(5) | 1404.40092(2) | 17(1) |
| 1392.97077(2) | 21(1) | 1404.40092(1) | 17.1(8) |
| 1392.97657(2) | 20(1) | 1404.41177(2) | 18(2) |
| 1392.98712(2) | 20(1) | 1404.43773(4) | 17(3) |
| 1392.99624(3) | 20(2) | 1404.56642(2) | 20(1) |
| 1393.26645(3) | 19(2) | 1404.58279(3) | 22(2) |
| 1393.27346(6) | 21(4) | 1405.5026(5) | 19(3) |
| 1393.36294(9) | 18(5) | 1405.54051(2) | 18(2) |
| 1393.39037(2) | 17.7(8) | 1406.10179(3) | 18(1) |